\documentclass[prb,twocolumn,showpacs,preprintnumbers,boldmath,amsmath,amssymb,floatfix]{revtex4-1}
\usepackage{times}
\usepackage{graphicx}
\usepackage{amssymb}
\usepackage{amsmath}
\usepackage{amsfonts}
\usepackage{dcolumn}
\usepackage{braket}
\usepackage{bm}
\usepackage{multirow}
\usepackage[colorlinks=true,
            linkcolor=red,
            urlcolor=blue,
            citecolor=blue]{hyperref}
\usepackage{amstext}
\usepackage{framed}
\begin{document}

\title{Local inversion symmetry breaking and spin-phonon coupling in perovskite GdCrO$_3$}

\author{Sudipta Mahana$^1$$^,$$^2$, Bipul Rakshit$^3$, Raktima Basu$^4$, Sandip Dhara$^4$, Boby Joseph$^5$, U. Manju$^6$, Subhendra D. Mahanti$^7$ and D. Topwal$^1$$^,$$^2$$^,$}

\email {dinesh.topwal@iopb.res.in, dinesh.topwal@gmail.com}

\affiliation{$^1$Institute of physics, Sachivalaya Marg, Bhubaneswar - 751005, India\\
$^2$Homi Bhabha National Institute, Training School Complex, Anushakti Nagar, Mumbai - 400085, India\\
$^3$Center for Superfunctional Materials, Ulsan National Institute of Science and Technology, Ulsan - 44919, South Korea\\
$^4$Surface and Nanoscience Division, Indira Gandhi Centre for Atomic Research, Kalpakkam - 603102, India\\
$^5$Elettra, Sincrotrone Trieste, Strada Statale 14, Km 163.5, Basovizza, Trieste - 34149, Italy\\
$^6$CSIR -Institute of Minerals and Materials Technology, Bhubaneswar - 751013, India\\
$^7$ Department of Physics and Astronomy, Michigan State University, East Lansing, Michigan 48824, USA}

\date{\today}

\begin{abstract}
Our detailed temperature dependent synchrotron powder x-ray diffraction studies along with first-principles density functional perturbation theory calculations, enable us to shed light on the origin of ferroelectricity in GdCrO$_3$. The actual lattice symmetry is found to be noncentrosymmetric orthorhombic $Pna2_1$ structure, supporting polar nature of the system. Polar distortion is driven by local symmetry breaking and by local distortions dominated by Gd off-centering. Our study reveals an intimate analogy between GdCrO$_3$ and YCrO$_3$. However, a distinctive difference exists that Gd is less displacive compared to Y, which results in an orthorhombic $Pna2_1$ structure in GdCrO$_3$ in contrast to monoclinic structure in YCrO$_3$ and consequently, decreases its polar property. This is due to the subtle forces involving Gd-4$f$ electrons either directly or indirectly. A strong magneto-electric coupling is revealed using Raman measurements based analysis in the system below Cr-ordering temperature, indicating their relevance to ferroelectric modulation.

\end{abstract}

\maketitle

\section{Introduction}
Multiferroics with simultaneous existence of ferroelectricity and (anti)ferromagnetism have been of great interest over the past several decades not only for their potential technological applications but also for fundamental understanding since these materials are anticipated to potentially bring about a very large magnetoelectric (ME) effect. In spite of their huge potentials, multiferroic materials are rare because of having contradictory requirements, as magnetism requires an odd number of $d$-electrons, while ferroelectricity generally occurs only in materials without $d$-electrons \cite{hill2000there}. Such contrary requirements has led to an upsurge in research activities in this field aimed at identifying alternative mechanisms by which these degrees of freedom can coexist and couple strongly. Probably the best known ferroelectrics are $AB$O$_3$ perovskite-type oxides such as BaTiO$_3$ \cite{cohen1992origin,kwei1993structures}, in which ferroelectricity originates from the off-centering of the Ti with respect to the oxygen octahedral cage due to the virtual hopping of electrons between empty Ti-$d$ and occupied O-$p$ states, whereas in BiFeO$_3$ \cite{wang2003epitaxial} and BiMnO$_3$ \cite{seshadri2001visualizing} ferroelectricity is Bi-6$s$ lone-pair driven, which results in the displacement of $A$-site ion from centrosymmetric positions with respect to the surrounding oxygen ions. Such materials are classified as proper ferroelectrics, where the origin of ferroelectricity is a structural instability towards the polar state associated with electronic pairing \cite{cheong2007multiferroics}. \\

    In contrast, there exists a large variety of improper ferroelectrics such as orthorhombic rare-earth manganites ($R$MnO$_3$, $R$ = Gd, Tb, Dy) \cite{cheong2007multiferroics,kenzelmann2005magnetic,kimura2005magnetoelectric} in which ferroelectricity arises due to the breaking of inversion symmetry from the spiral spin-order. The underlying mechanisms for the generation of electric polarizations are inverse Dzyaloshinskii-Moriya interaction, where the spin configuration displaces oxygen (ligand) ions through the electron-lattice interaction or spin-current model \cite{katsura2005spin}. Another mechanism that can lead to ferroelectricity is charge ordering, in which $B$-sites contain transition metal ions with different valancy: for example $R$NiO$_3$ \cite{giovannetti2009multiferroicity}. In hexagonal rare-earth manganites ($R$MnO$_3$, $R$ = Ho-Lu, Y) polarization arises from the tilting of MnO$_5$ polyhedra accompanied by displacement of the $R$ ions, therefore they have been coined as improper geometric ferroelectrics \cite{cheong2007multiferroics,van2004origin}. Apart from these, a low-temperature ferroelectric phase is observed in Gd(Dy)FeO$_3$,  which arises due to the exchange striction between Gd(Dy) and Fe spins \cite{tokunaga2009composite,tokunaga2008magnetic}, whereas inverse Dzyaloshinskii-Moriya interaction mechanism makes SmFeO$_3$ a room temperature ferroelectric \cite{lee2011spin}. Perovskite CdTiO$_3$ is a unique system, in which ferroelectricity is driven by a phase transition from the centrosymmetric orthorhombic structure ($Pbnm$) to a non-centrosymmetric structure ($Pna2_1$) via displacement of Ti and O ions, even though overall orthorhombic symmetry is maintained \cite{shan2003ferroelectric,moriwake2011first}. Recently, a new mechanism is proposed where the rotation of O octahedra coupled with lattice distortion leads to a ferroelectric phase \cite{benedek2011hybrid,benedek2012polar,young2015anharmonic}.\\
   Moreover, there are diverging opinions about the existence and/or origin of ferroelectricity in rare-earth orthochromites ($R$CrO$_3$). Most of the members of rare-earth orthochromites ($R$CrO$_3$) family have been predicted to be biferroic with a reasonably high ferroelectric (FE) transition temperature (above magnetic transition temperature), caused by polar movement of  $R$-ions associated with phonon instability at zone-centre similar to other perovskite ferroelectrics like PbTiO$_3$  \cite{serrao2005biferroic, ray2008coupling,sahu2007rare,waghmare1997ab}. From neutron pair distribution function (PDF) analysis, YCrO$_3$ has been reported to possess a locally non-centrosymmetric monoclinic structure ($P2_1$) via Cr off-centering in the ferroelectric state though the average crystal structure is centrosymmetric ($Pnma$/$Pbnm$) \cite{ramesha2007observation}. The structural instability was also supported from theoretical calculation, however this polar instability mode is associated with Y displacements in a direction opposite to
that of the oxygen cage and Cr atom \cite{serrao2005biferroic,ray2008coupling}. It suggests that ‘local non-centrosymmetry’ could play an important role in understanding the ferroelectric properties in the family of $R$CrO$_3$ \cite{rajeswaran2012field}. \\

In addition to the structural distortion ideas where the magnetic coupling plays a minor role; several studies have reported that the multiferroicity in $R$CrO$_3$ is due to strong interaction between magnetic $R$ and weakly coupled ferromagnetic (canted) Cr ions below the magnetic-ordering temperature of Cr(T$_N$) along with a lower symmetry structure than $Pbnm $ \cite{rajeswaran2012field,apostolov2015microscopic}. Also, Raman studies show anomalous change of phonon frequency and a decrease in phonon life times across the mutiferroic transition temperature, both in the modes involving CrO$_6$ octahedra and magnetic $R$-ion \cite{bhadram2013spin}. This suggests that ferroelectricity in $R$CrO$_3$ is driven by magnetostriction mechanism caused by 3$d$-4$f$ coupling, resulting in the displacement of the $R$-ion and the octahedral distortion via oxygen displacements. This suggests that one has to have a magnetic $R$-ion in order to stabilize a ferroelectric state in RCrO$_3$. Recently, some $R$CrO$_3$ ($R$ = Sm and Ho) compounds are suggested to have structural transformation from centrosymmetric $Pbnm$ to non-centrosymmetric $Pna2_1$ sub-group, which is responsible for the polar order \cite{ghosh2014polar,ghosh2015atypical}. In these systems magnetic coupling between $R$-ion and the matrix is not important in stabilizing a ferroelectric phase as it develops in the paramagnetic state. On the contrary, recently it is reported that GdCrO$_3$ can only be ferroelectric at very low temperature via magnetostriction effect and it is necessary to have Gd${^+}$-Cr${^+}$ interaction and $G$-type magnetic structure in both to break the inversion-symmetry as driven by antipolar $X$-mode instability \cite{zhao2017improper}. Considering all the contradicting possibilities, it is important to understand the microscopic origin and mechanism of ferroelectricity in GdCrO$_3$ at relatively high temperatures. \\
In this manuscript we discuss the results of detailed studies on the possible non-centrosymmetric structure of GdCrO$_3$ using X-ray diffraction (XRD) measurements and from first principle calculations. We also investigate the possible polar phonon mode instability in its cubic structure as it plays a crucial role to understand classic ferroelectrics. Further we also discuss the magnetoelectric coupling in the material from temperature dependent Raman measurements.

\section{Experimental and Theoretical Methods}

GdCrO$_3$ sample was prepared by the solid state reaction method as reported elsewhere \cite{mahana2016giant, mahana2017complex}. 
 Phase purity of the sample was confirmed by
powder x-ray diffraction (XRD) measurements carried out in D8 advanced diffractometer equipped with Cu $K$$\alpha$ radiation.
Temperature dependent XRD measurements were performed at the XRD1 beamline at ELETTRA synchrotron radiation facility using photons of wavelength 0.85507 ${\AA}$. Reitveld refinements of the diffraction patterns were performed using the ${Fullprof }$ package.  The vibrational properties of the sample were measured using a micro-Raman spectrometer (inVia, Renishaw, UK) with 514.5 nm excitation of an Ar$^+$ laser. Spectra were collected in the back scattering configuration using a thermoelectrically cooled CCD camera as the detector. A long working distance 50X objective with numerical aperture of 0.45 was used for the spectral acquisition. In order to carry out the temperature dependent Raman spectroscopic measurements, the sample was kept in a Linkam (THMS600) stage, driven by an auto-controlled thermoelectric heating and cooling function within a temperature range of 80 to 300 K.\\

Our theoretical calculations of the structural properties were based on density functional theory, using generalized gradient approximation (GGA) with Perdew Burke Ernzerhof \cite{perdew1996generalized} parameterization for the exchange correlation potential, the projector argumented wave (PAW) method\cite{kresse1999ultrasoft}, and a plane-wave basis set, as implemented in the Vienna ab-initio simulation package (VASP) \cite{kresse1996efficient}
The interaction between ions and electrons is approximated with ultra-soft pseudo-potentials, treating 3$d$ and 4$s$ for Cr; 2$s$ and 2$p$ electrons for O; 5$s$, 5$p$, 5$d$ and 6$s$ for Gd as valence electrons considering the 4$f$ electrons inside the ionic core. For Brillouin zone sampling, we chose 12{$\times$}12{$\times$}8 and 6{$\times$}6{$\times$}6 Monkhorst-Pack $k$-point mesh \cite{monkhorst1976special} for orthorhombic and cubic phase, respectively and the wave-function was expanded in a basis set consisting of plane waves with kinetic energies less than or equal to 770 eV. Using these parameters, an energy convergence of less than 1 meV/formula unit (f.u.) was achieved. Structures were relaxed until residual Hellmann–Feynman (HF) forces were smaller than 0.001 eV/$\AA{}$ while maintaining the symmetry constraints of the given space group. In order to impose G-type antiferromagnetic ordering in cubic structure, the unit cell was doubled along $<$111$>$ direction, which resulted in 10 atoms unit cell \cite{ray2008coupling}. Experimental values of the volume of the unit cell was used for the calculation since ferroelectricity is very sensitive to the values of lattice parameters \cite{serrao2005biferroic, ray2008coupling,looby1954yttrium}. The phonon frequencies were calculated in high symmetry directions using the 2{$\times$}2{$\times$}2 supercell. The real-space force constants of the supercell were calculated using VASP via density functional perturbation theory (DFPT). The unit cell results in 30 phonon branches: 3 acoustic which have a zero frequency at $k$ = (0,0,0) and 27 optical, some of which are triply degenerate. We are mainly interested in optical modes with imaginary phonon frequencies corresponding to instabilities in the structure. Due to the doubling of the unit cell along the $<$111$>$ direction, we could access zone boundary phonon modes at the $R$- point along with the zone-center modes at $\Gamma$ point \cite{ray2008coupling}. These phonon calculations were done considering Gd-4$f$ electrons as both valence state and core state in order to find out role of $f$-electrons on the ferroelectric behavior in the system. G-type magnetic structure was considered for both Cr and Gd moments in the calculation \cite{zhao2017improper,cooke1974magnetic}. The electric polarization was calculated using the Berry phase method \cite{king1993theory}, as implemented in VASP.  The utility tool phonopy \cite{togo2008first} was used to obtain phonon frequencies and phonon dispersions over the entire Brillouin zone. 

\section{Results and discussion}

GdCrO$_3$ crystallizes in perovskite structure with Goldschimdt’s tolerance factor, $t$ = $\frac{(r_{{Gd}^{3+}} + r_{O^{2-}})}{\sqrt2(r_{{Cr}^{3+}} + r_{O^{2-}})}$ =  0.862, indicating an orthorhombically distorted structure \cite{coey2009mixed,hines1997atomistic}. FIG. 1(a) and (b) depict synchrotron x-ray diffraction patterns acquired at 300 K and 100 K respectively along with the corresponding Rietveld refined data using $Pbnm$ space group superimposed on it. Reasonably small values of reliability parameters (For 300 K: $R_w$ $\sim$ 0.084, $R_{exp}$ $\sim$ 0.041 and $\chi^2$ $\sim$ 4.16, while for 100 K: $R_w$ $\sim$ 0.088, $R_{exp}$ $\sim$ 0.043 and $\chi^2$ $\sim$ 4.23 ) indicate good quality of the fitting and suggest that the centrosymmetric $Pbnm$ space group persists in the entire (studied) temperature range. Additionally, the compound undergoes a G-type magnetic ordering in Cr-sublattice below T$_N$ = 169 K \cite{cooke1974magnetic, rajeswaran2012field} (whereas Gd-sublattice remains in paramagnetic state). A weak electric polarization is also observed at the same temperature suggesting a magneto-electric coupling \cite{rajeswaran2012field}. It has to be noted that globally centrosymmetric magnetic and crystal structures are not compatible with the observation of ferroelectric phase. It strongly indicates a possibility of local non-centrosymmetric structure in GdCrO$_3$, which may be responsible for its ferroelectric property. \\
 
\begin{figure}[!ht]
 \centering
 \includegraphics[height=10cm,width=7cm]{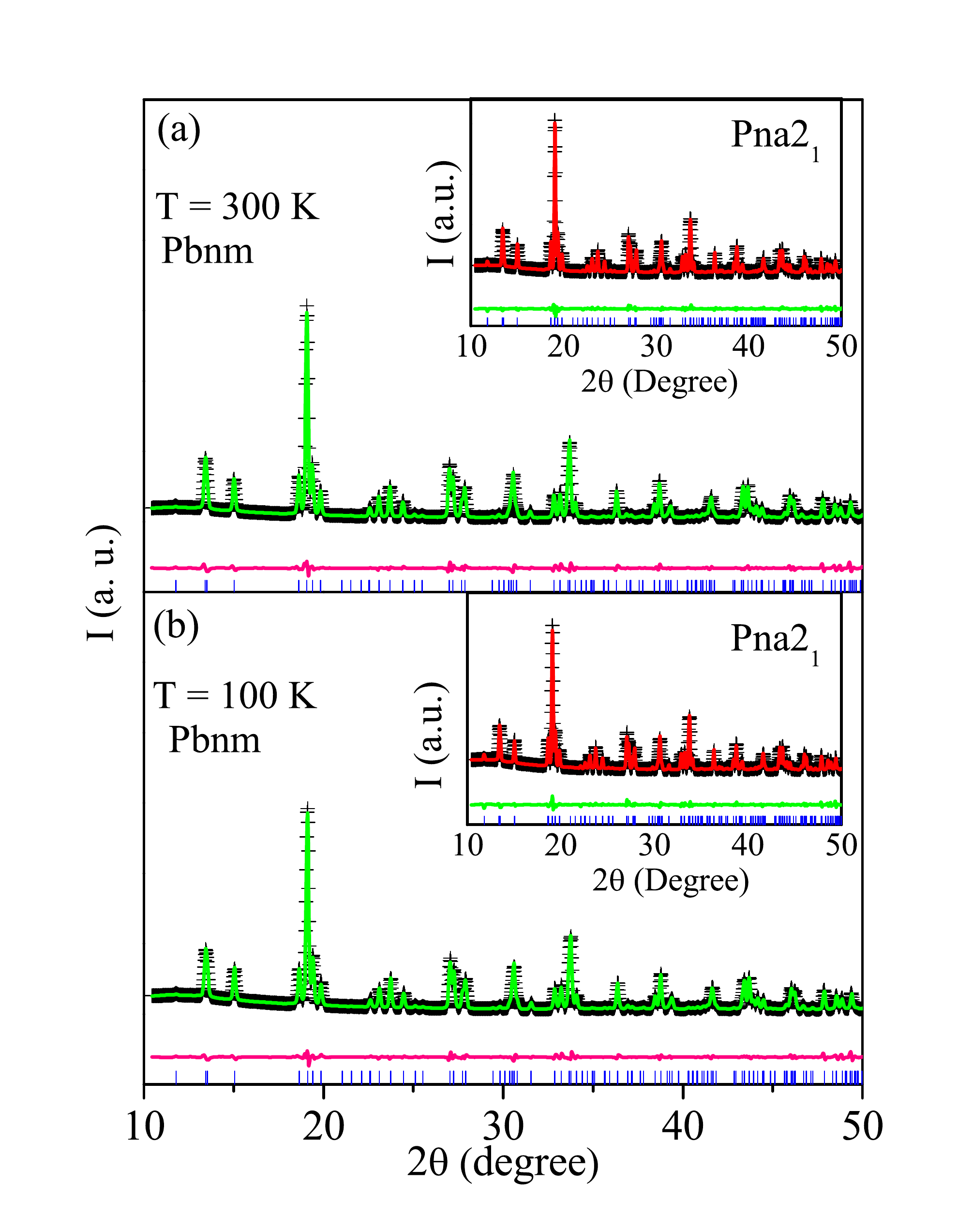}
 \caption{X-ray powder diffraction patterns (symbol) obtained at 300 K (a) and 100 K (b) with the refinement patterns (continuous curve) using ${Pbnm}$ space group superimposed on it. Insets represent the same with Reitveld refinement using ${Pna2_1}$ space group.}
 \end{figure}

Recent synchrotron x-ray diffraction studies on the other members of rare-earth chromite family ($R$CrO$_3$, $R$ = Sm, Ho and Nd) over a wide temperature range reveals a structural transition from a high temperature centrosymmetric $Pbnm$ space group to low temperature non-centrosymmetric $Pna2_1$ sub-group close to the onset of polar order \cite{ghosh2014polar,ghosh2015atypical,indra2016erratum}. Also, it has been proposed that local noncentrosymmetry drives ferroelectricity in YCrO$_3$, though, its average crystal structure is centrosymmetric \cite{serrao2005biferroic,ramesha2007observation}. Keeping this in mind XRD patterns of GdCrO$_3$ were also refined with non-centrosymmetric $Pna2_1$ space group, as depicted in insets of FIG. 1 and the reliability parameters were surprisingly very similar to the $Pbnm$ space group for both 300 K and 100 K suggesting that the average long-range ordering as depicted by x-ray diffraction can not alone provide an answer to the origin of ferroelectricity in GdCrO$_3$.\\

We hence performed first-principles density functional perturbation theory calculations using experimental lattice parameters, optimizing only the internal co-ordinates. An energy lowering of 1.7 meV/f.u was observed for $Pna2_1$ structure as compared to the $Pbnm$ structure; significantly less compared to YCrO$_3$ which has a non-cenrosymmetric monoclinic ($P2_1$) phase \cite{serrao2005biferroic} and quite large compared to CdTiO$_3$ which has a non-cenrosymmetric orthorhombic ($Pna2_1$) structure \cite{moriwake2011first}.  The room temperature experimental lattice parameters for $Pbnm$ and  $Pna2_1$ space group are summarized in Table I. Based on the above observations, we believe that the ground state structure of GdCrO$_3$ is non-centrosymmetric $Pna2_1$, as it favors non-zero polarization. However, the distortion in the structure might be very small, hence it is not distinguished by XRD measurements. Recent theoretical calculations allowing for magnetic interaction for different crystal symmetries by Zhao ${et}$ ${al.}$ suggested that only $Pna2_1$ crystal symmetry gives a sizable polarization for GdCrO$_3$, which is in agreement with our results \cite{zhao2017improper}. However, according to them GdCrO$_3$ can only be ferroelectric at very low temperature as it is necessary to have magnetic ordering in Gd-sublattice (G-type) along with that of Cr-sublattice to break the inversion symmetry $via$ exchange-striction mechanism. They also found that the distortion was associated with cubic structure. In contrast we find a stable non-centrosymmetric $Pna2_1$ structure even in the absence of any magnetic coupling between Cr and Gd moments and induced magnetic order of the Gd moments.\\

\begin{figure}[!ht]
 \centering
 \includegraphics[height=7.8cm,width=9.5cm]{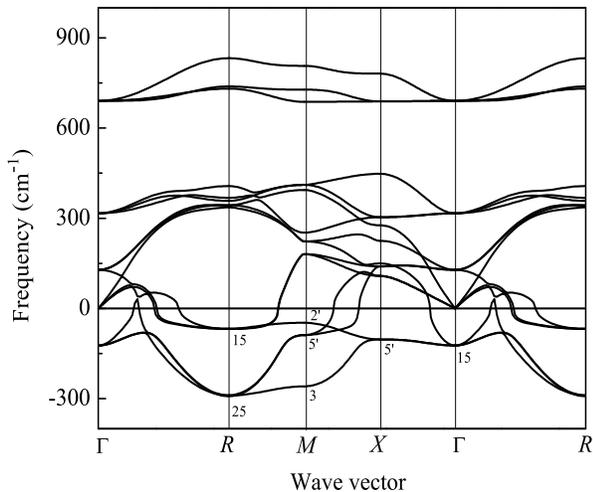}
 \caption{Phonon dispersion curves for the cubic phase of GdCrO$_3$. The labels indicate the symmetry of unstable modes.}
 \end{figure}

To understand the origin of ferroelectricity in GdCrO$_3$ at relatively high temperature we performed phonon calculations for the cubic perovskite structure of GdCrO$_3$, to access various structural instabilities in the system. Structural instability studies have been used to examine a large variety of ferroelectric perovskite oxides \cite{serrao2005biferroic, ray2008coupling,sahu2007rare,waghmare1997ab,ghosez1999lattice,bhattacharjee2009engineering} . Calculated phonon dispersion curves are shown in FIG. 2, with imaginary frequencies plotted on the negative axis. Soft modes occur over a wide range of wave vectors, with strong instabilities at $R$ ($R_{25}$) and $M$ ($M_3$), symmetry points which are associated with the octahedral rotations. Simultaneous condensation of these soft modes result in cubic-to-orthorhombic phase transition. Polar mode instabilities at $R$, $X$ and $\Gamma$ points, are associated with displacements of Gd and oxygen atoms, case similar to that for YCrO$_3$ \cite{serrao2005biferroic}. Modes at $R$ ($R_{15}$) and $X$($X_{5'}$) points correspond to anti-ferroelectric distortions, and can only give rise to non-zero polarization below Gd-ordering temperature via magneto-striction effect with the dominant contribution from $X$ mode as reported by Zhao ${et}$ ${al.}$\cite{zhao2017improper}. The $\Gamma$ ($\Gamma_{15}$) mode on the other hand is responsible for the polarization at relatively high temperature as described in YCrO$_3$ and other ferroelectric perovskite compounds \cite{serrao2005biferroic, ray2008coupling,waghmare1997ab}. \\

We also determined phonon frequencies at the $\Gamma$ point, which correspond to phonons at $\Gamma$ and $R$ points of the primitive unit cell. Treating Gd 4$f$ as core state, we found two triply degenerate zone-center instabilities at 144$i$ cm$^{-1}$ ($\Gamma_{25}$) and 149$i$ cm$^{-1}$ ($\Gamma_{15}$) and two triply degenerate zone-boundary instabilities ($R$) at 78$i$ cm$^{-1}$ ($R_{15}$),  and 335$i$ cm$^{-1}$ ($R_{25}$) similar to YCrO$_3$ and other $d^3$ systems \cite{ray2008coupling}. By considering Gd 4$f$ state as the valence state in the calculations, we noticed $\Gamma_{25}$ is no longer unstable. The other three negative modes are present there, however, their magnitudes are quite less such as 269$i$ cm$^{-1}$ ($R_{25}$), 91$i$ cm$^{-1}$ ($\Gamma_{15}$) and 34$i$ cm$^{-1}$ ($R_{15}$). This clearly indicates a strong influence of Gd-4$f$ electrons on the various instability modes of the cubic structure.
 
\begin{figure}[!ht]
 \centering
 \includegraphics[height=7.8cm,width=6cm]{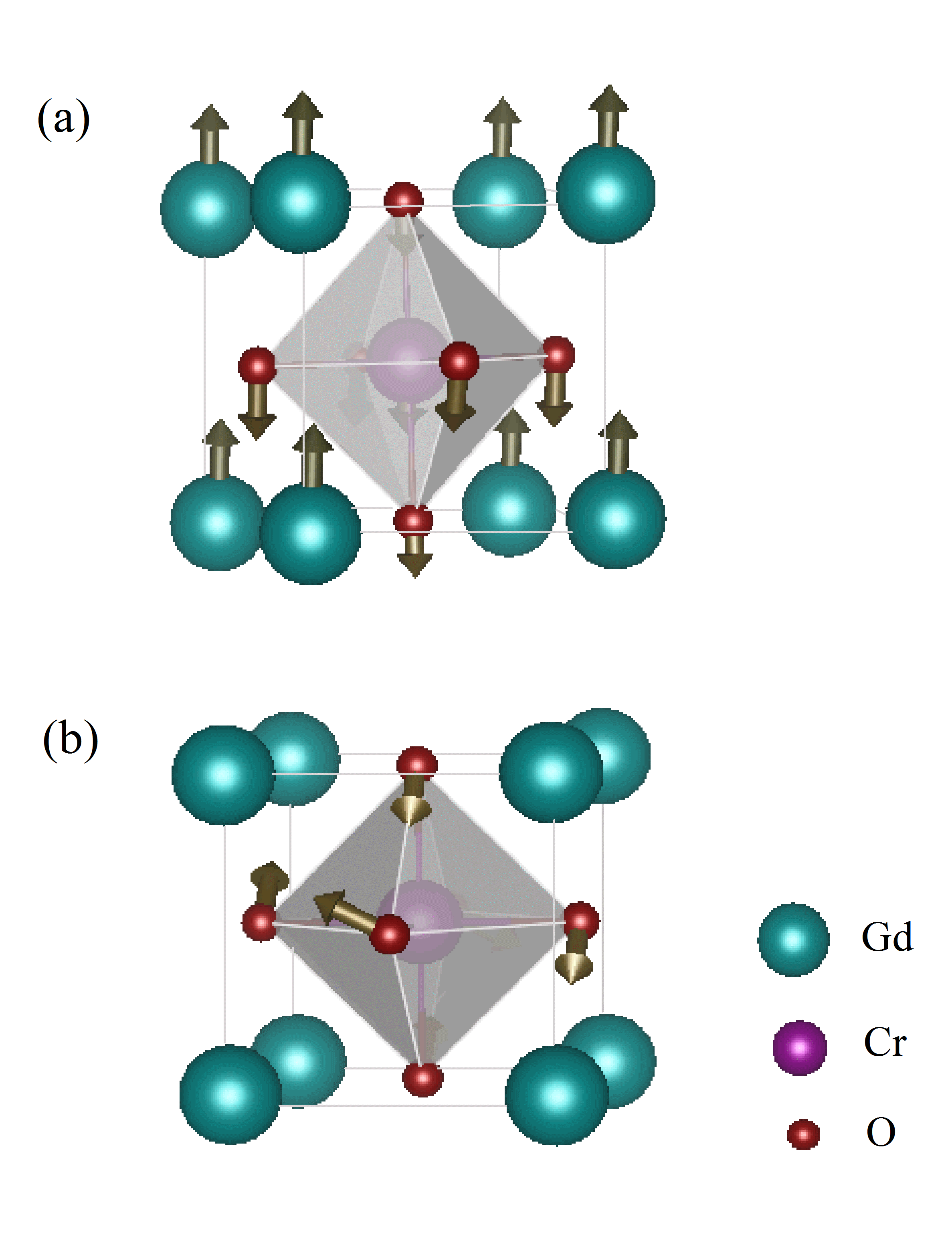}
 \caption{Visualization of eigenvectors of unstable (a) polar $\Gamma_{15}$ and (b) antiferrodistortive $R_{25}$ modes.}
 \end{figure}

 The weakest instability mode, $R_{15}$ involves the displacement of the Gd cations along with small oxygen displacements and these are antiparallel in neighboring unit cells. The next instability $\Gamma_{25}$ mode involves oxygen displacements only.  The $\Gamma_{15}$ mode (FIG. 3a) involves mainly the Gd ion movement in a direction opposite to that of the oxygen cage and Cr ions where as Cr and O ions move in the same direction, resulting in a ferroelectric polar structural distortion very similar to YCrO$_3$ \cite{serrao2005biferroic,ray2008coupling}. The strongest instability $R_{25}$ (FIG. 3b) is a anti-ferrodistortive mode corresponding to the rotation of the corner connected oxygen octahedra. To probe the strength of ferroelectric instability, we displaced the atoms toward the eigenvectors for polar $\Gamma_{15}$ mode and relaxed it, which resulted in an energy lowering by 21.2 meV/f.u., slightly less compared to YCrO$_3$ \cite{serrao2005biferroic} and other perovskite ferroelectrics like PbTi$O_3$ \cite{waghmare1997ab}. 

We calculated the magnitude of polarization in Berry phase method  and found it to be 0.56 $\mu$C/cm$^2$, this is of the same order of magnitude as the experimental value (0.7 $\mu$C/cm$^2$) \cite{rajeswaran2012field}. It is surprising that polarization value in GdCrO$_3$  is one order of magnitude less than that of YCrO$_3$ \cite{serrao2005biferroic} even though Born effective charges (BEC) of Gd is quite large compared to Y as given in Table II. However, only BEC can not define the tendency of a certain material towards ferroelectricity \cite{ederer2011mechanism}. Larger BEC in Gd as compared to Y may possibly be due to the magnetic character of Gd. Also the ionic radii of Gd$^{3+}$ (1.08 $\AA{}$) and Y$^{3+}$ (1.04 $\AA{}$) ions are very similar. The difference in polarizations arises from smaller displacement of  Gd compared to Y and is due to the subtle forces involving Gd 4$f$ electrons either directly or indirectly.  In addition, larger BEC of Cr and one of the oxygen atoms in GdCrO$_3$ compared to that of YCrO$_3$ is due to its stronger Cr-O-Cr superexchange interaction; consequently, which leads to a relatively higher magnetic transition temperature in GdCrO$_3$ than that of YCrO$_3$ \cite{ray2008coupling}.  \\

Ferroelectric behavior in GdCrO$_3$ is based on the competition between polar mode and antiferrodistortive rotational mode \cite{zhong1995competing,aschauer2014competition}. The polar mode favors non-centrosymmetric ferroelectric phase where as  antiferrodistortive instability (octahedral rotation) hinders the ferroelectric ordering by inducing $R$-site antipolar displacements and leads to centrosymmetric phase. The competition among these two modes result in various structures, such as rhombohedral, which is associated with only ferroelectric instability. Successive increase of antiferrodistortive instability combined with decrease of polar instability results into monoclinic, tetragonal and orthorhombic phases, respectively, which in turn suppress the ferroelectric property gradually in the structure \cite{zhong1995competing}. It can be noted that both polar and antiferrodistortive instability modes are weaker in GdCrO$_3$ than in YCrO$_3$ \cite{ray2008coupling}. The dominant contribution of  antiferrodistortive mode together with ferroelectric mode stabilizes $Pna2_1$ phase in GdCrO$_3$ and decrease the ferroelectric property compared to YCrO$_3$, which stabilizes in monoclinic $P2_1$ phase \cite{ramesha2007observation}. The intrinsic differences in the bonding in monoclinic YCrO$_3$ and orthorhombic GdCrO$_3$ leads to different magnitude of polarization.

\begin{table} 
\caption{ Experimental lattice parameters of GdCrO$_3$ both in $Pbnm$ and $Pna2_1$ structure at 300 K .}

\begin{tabular} {cccccccccccccccc} \hline

                      Lattice   &&& $Pbnm$       &&& $Pna2_1$ \\
               parameter &&&  (\AA{})     &&& (\AA{})    \\ \hline
                a         &&& 5.306  &&& 5.519  & &  \\
                b         &&& 5.513   &&& 5.309 & &   \\
                c         &&& 7.594   &&& 7.590 & &  \\
 & & & \\ \hline

\end{tabular}
\end{table}

\begin{table} 
\caption{ The XX component of Born Effective charge tensor for cubic GdCrO$_3$ and YCrO$_3$ compounds..}

\begin{tabular} {cccccccccccccccc} \hline

Compound  &&& Z$_R^\star$  &&& Z$_{Cr}^\star$        &&& Z$_{O_x}^\star$  &&& Z$_{O_{y,z}}^\star$ \\
                                &&&   \\ \hline

GdCrO$_3$$^a$         &&&   4.75        &&& 3.51   &&& -3.84 && & -2.19 \\                                                                                                                                                                                                            
 & & & \\ \hline

YCrO$_3$$^b$       &&&   4.45         &&& 3.44   &&& -2.62 && & -2.66 \\
             
&&&\\ \hline
$^a$This work and  $^b$Ref.\cite{ray2008coupling}.

\end{tabular}
\end{table}

As discussed previously XRD and theoretical studies reveal the probable non-centrosymmetric $Pna2_1$ crystal structure in ferroelectric state for GdCrO$_3$. Displacements of Gd-atoms combined with antiferrodistortive distortion of octahedra via movements of specific oxygen ions lift certain symmetries of centrosymmetric $Pbnm$ structure and stabilize the lower symmetry $Pna2_1$ structure. In short, there is a coupling and competition between antiferrodistortive zone-boundary and polar zone-center instabilities and consequent structural rearrangements are responsible for the emergence of spontaneous polarization. \\

\begin{figure}[!ht]
 \centering
 \includegraphics[height=10cm,width=8cm]{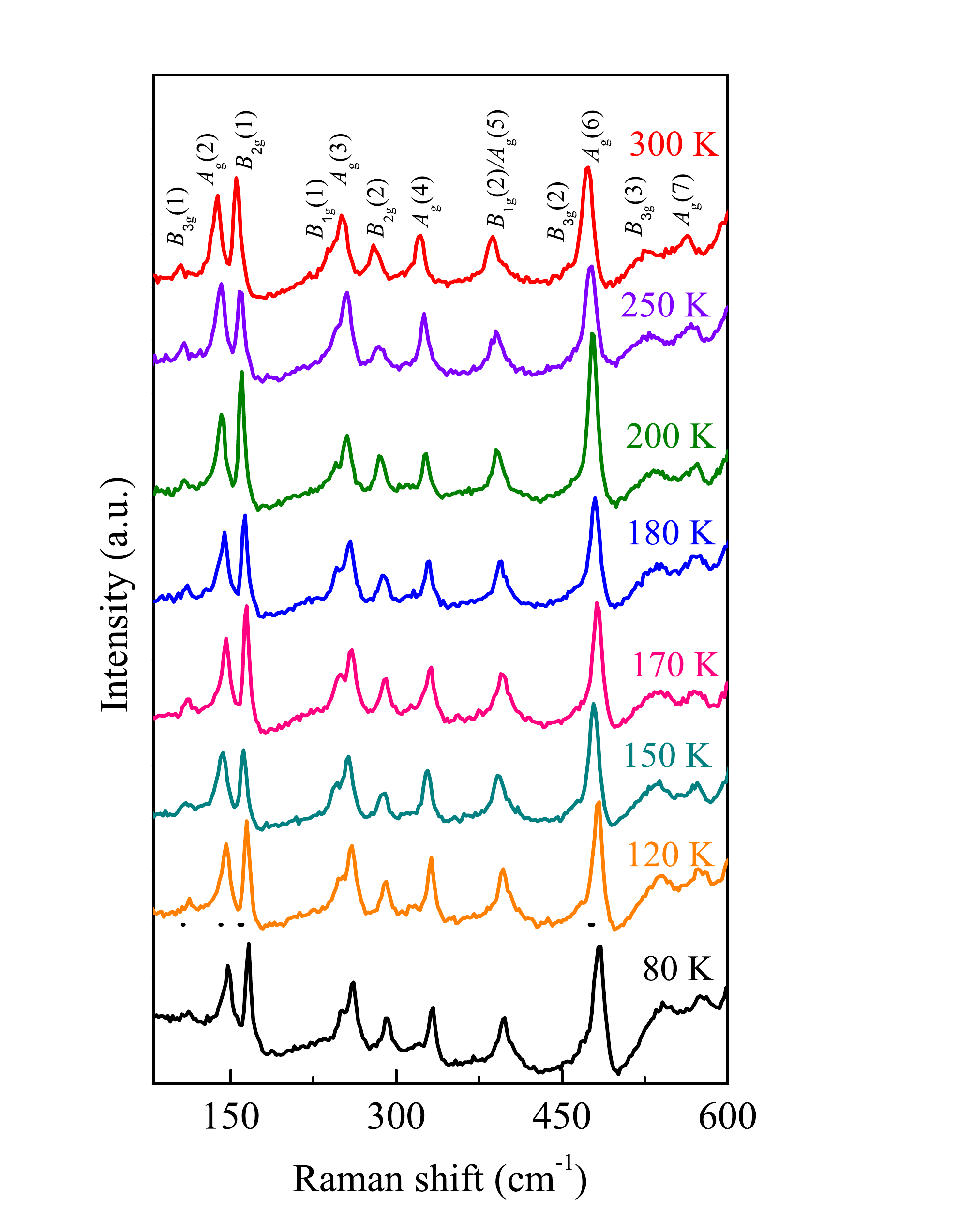}
 \caption{Raman spectra of GdCrO$_3$ at few selective temperatures both above and below the magnetic/ferroelectric ordering temperature (169 K).}
 \end{figure}

In addition to the structural studies through X-ray diffraction, we also performed temperature dependence Raman spectroscopic studies, as it is a powerful and sensitive technique for detecting more subtle structural rearrangements and microscopic changes across the phase transitions such as evolution of phonons, magnons and electromagnons in multiferroic materials. Based on group theoretical analysis 24 first order Raman active modes are expected for GdCrO$_3$ which are classified as $\Gamma_{Raman}$ = 7$A$$_g$ +5$B$$_{1g}$ +7$B$$_{2g}$ +5$B$$_{3g}$, involving vibration of Gd and oxygen atoms \cite{udagawa1975influence}.  FIG. 4 depicts the temperature dependent Raman spectra at a few selected temperatures both below and above the transition temperature. We see only 14 Raman active modes. The absence of other predicted modes are due to very low intensity, which are below the detection limit of the instrument or beyond our experimental range. The phonon modes below 200 cm$^{-1}$ arise from the movement of Gd-atoms, $B_{1g}$(1) and $A_g$(3) involving out-of-phase and in-phase octahedral $z$-rotations, respectively, B$_{2g}$(2) and A$_g$(4) are related to Gd-O vibrations in GdO$_{12}$ polyhedra, B$_{1g}$(2)/A$_g$(5) involve the out-of-phase/in-phase octahedral $y$-rotations, B$_{3g}$ (2) is associated with out-of- phase bending,  A$_{g}$ (6)  involves octahedral bending modes, B$_{3g}$(3) is associated with in-phase O2 scissor-like vibration and A$_g$(7) arises from antisymmetric stretching vibration of octahedra \cite{bhadram2013spin,weber2012phonon,bhadram2014effect,udagawa1975influence}. \\

 At first glance, no new Raman active modes emerge down to 80 K from 300 K. Further, to examine subtle structural changes and the presence of any interactions between lattice and magnetic degrees of freedom, i.e., spin-phonon coupling, the Raman spectra were analyzed by Lorentzian fitting of the peaks. The intrinsic anharmonic contribution to temperature variation of phonon frequency of Raman modes can be explained by the following relation \cite{balkanski1983anharmonic},
 \begin{eqnarray}
\omega_{anh}(T) &=& \omega(0) - A[{1+\frac{2}{e^\frac{\hbar\omega(0)}{2k_BT}-1}}]\nonumber \\  &&
- B[{1+\frac{3}{e^\frac{\hbar\omega(0)}{3k_BT}-1}} +\frac{3}{(e^\frac{\hbar\omega(0)}{3k_BT}-1)^2}],
\end{eqnarray}
 where $\omega$(0) is zero-Kelvin frequency of the mode in harmonic approximation, T is in K, A and B are anharmonicity coefficients for cubic and quartic anharmonic processes, respectively. FIG. 5(a) represents the temperature evolution of octahedral rotational mode around y-axis ($B_{1g}$(2)/$A_g$(5) represented as circles, along with their fitting using equation (1) marked as dotted line. Below the transition temperature, it shows a pronounced softening from the intrinsic anharmonic contribution. The anomalous behavior of these phonon modes across $T_N$ can be explained by exchange-striction effect. To understand more about the origin of anomalous behavior of various phonon modes such as the presence of spin-phonon coupling, it is necessary to study the temperature dependence of corresponding linewidths, as Raman linewidths are related to the phonon lifetime which will not be affected by subtle volume changes due to exchange-striction effect/magnetoelastic coupling. \\

\begin{figure}[!ht]
 \centering
 \includegraphics[height=10cm,width=8cm]{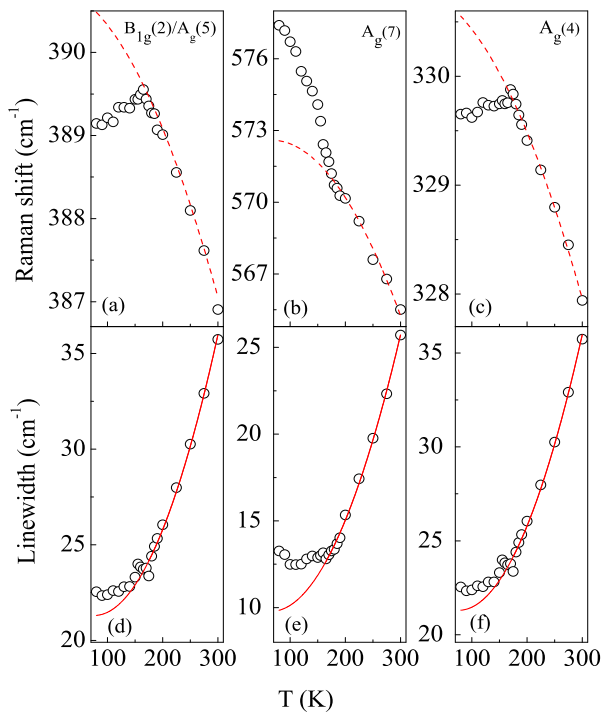}
\caption{(a)-(c) represent temperature dependence of frequencies of few selective modes (octahedral rotation with respect to $y$-axis ($B_{1g}$(2)/($A_{g}$(5) ), antisymmetric stretching ($A_g$(7)) and Gd-O vibration ($A_g$(4)), respectively. The dotted lines represent the fitted curves for anharmonic contributions to these modes according to Eq. (1). (d)-(e) represent line widths of corresponding modes and solid lines represent the fitted curves for anharmonic contributions according Eq. (1).}
 \end{figure}

FIG. 5(d) shows the temperature evolution of the linewidth of the mode related to octahedral rotations. The anomalies in the linewidth across the transition indicates the spin-phonon coupling in GdCrO$_3$. Such type of spin-phonon coupling was not observed in $R$CrO$_3$ having non-magnetic $R^{3+}$ ions such as Y and Lu etc \cite{bhadram2013spin}. This suggests the presence of spin-phonon coupling due to magnetic interaction between Gd$^{3+}$ and Cr$^{3+}$ moments which is mediated by the weak ferromagnetic coupling (canted)  of Cr-sublattice. The temperature variations of frequencies and linewidths of antisymmetric stretching mode ($A_g$(7)) are shown in FIG 5(b) and (e), respectively. Such hardening behavior of  the antisymmetric stretching mode in YCrO$_3$ has been explained by exchange-striction effect with a major contribution of 30-40 percent and the remaining contributions coming from magnetic contributions (Cr$^{3+}$-Cr$^{3+}$ interaction) \cite{udagawa1975influence}. Thus spin-phonon coupling can not be ignored in YCrO$_3$ although no significant anomaly in linewidth has been seen by Bhadram ${et}$ ${al.}$ in this system \cite{bhadram2013spin}. Similarly, in GdCrO$_3$ hardening behavior of  the antisymmetric stretching mode can be explained by exchange striction effect consistent with the reduction of unit cell volume. In addition, a pronounced anomaly is observed in the linewidth. This indicates a strong spin-phonon coupling, which can be explained by  the Gd$^{3+}$-Cr$^{3+}$ interaction in addition to a contribution from Cr$^{3+}$-Cr$^{3+}$ interaction. Sharma ${et}$ ${al.}$ observed considerable softening of the bending mode along with anomaly in its linewidth in YCrO$_3$ (non-magnetic $R$-ion) \cite{sharma2014phonons}, implying significant contribution of Cr$^{3+}$-Cr$^{3+}$ magnetic interaction to the spin-phonon coupling below the magnetic transition in YCrO$_3$. Moreover, lattice modes related to Gd atoms also show strong softening below the transition along with anomalies in their linewidths as clearly seen in $A_g$(4) mode (FIG. 5(c) and (f)). This suggests a possible displacement of Gd$^{3+}$ ion induced by strong spin-phonon coupling caused by Gd$^{3+}$-Cr$^{3+}$ interaction \cite{bhadram2013spin}. \\

 As discussed above, the anomalous behavior of various modes is mainly due to exchange-striction effect (lattice contribution) and spin-phonon coupling induced by Cr$^{3+}$-Cr$^{3+}$ and Gd$^{3+}$-Cr$^{3+}$ interactions right below the magnetic transition . 
  Granado ${et}$ ${al.}$  proposed that the spin-phonon coupling strength can be estimated for a given mode by relating the deviation of Raman mode frequency from intrinsic anharmonic contribution to the nearest neighbor spin-spin correlation function ($S_i$.$S_j$) as given by \cite{granado1999magnetic},
 \begin{eqnarray}
\Delta\omega_{sp-ph} = \lambda <S_i.S_j>,
 \end{eqnarray}
where $\lambda$ is the spin-phonon coupling coefficient. In molecular field approximation, the spin-spin correlation function can be described by the square of the sublattice magnetization \cite{iliev2007raman} and also by the normalized order parameter \cite{el2014local}. The temperature dependence 
of the frequency mode can be written as follows,
\begin{eqnarray}
\Delta\omega_{sp-ph} = \lambda S^2[1-(\frac{T}{T_N})^\gamma] \approx \lambda (\frac{M(T)}{M_{max}})^2,
 \end{eqnarray}
where  $T_N$ is Cr-ordering temperature, $S$ = 3/2 is the spin quantum number of Cr$^{3+}$ ion, $\gamma$ is the critical exponent, $M(T)$ is the magnetization as function of temperature ($T$).

\begin{figure}[!ht]
 \centering
 \includegraphics[height=8cm,width=7cm]{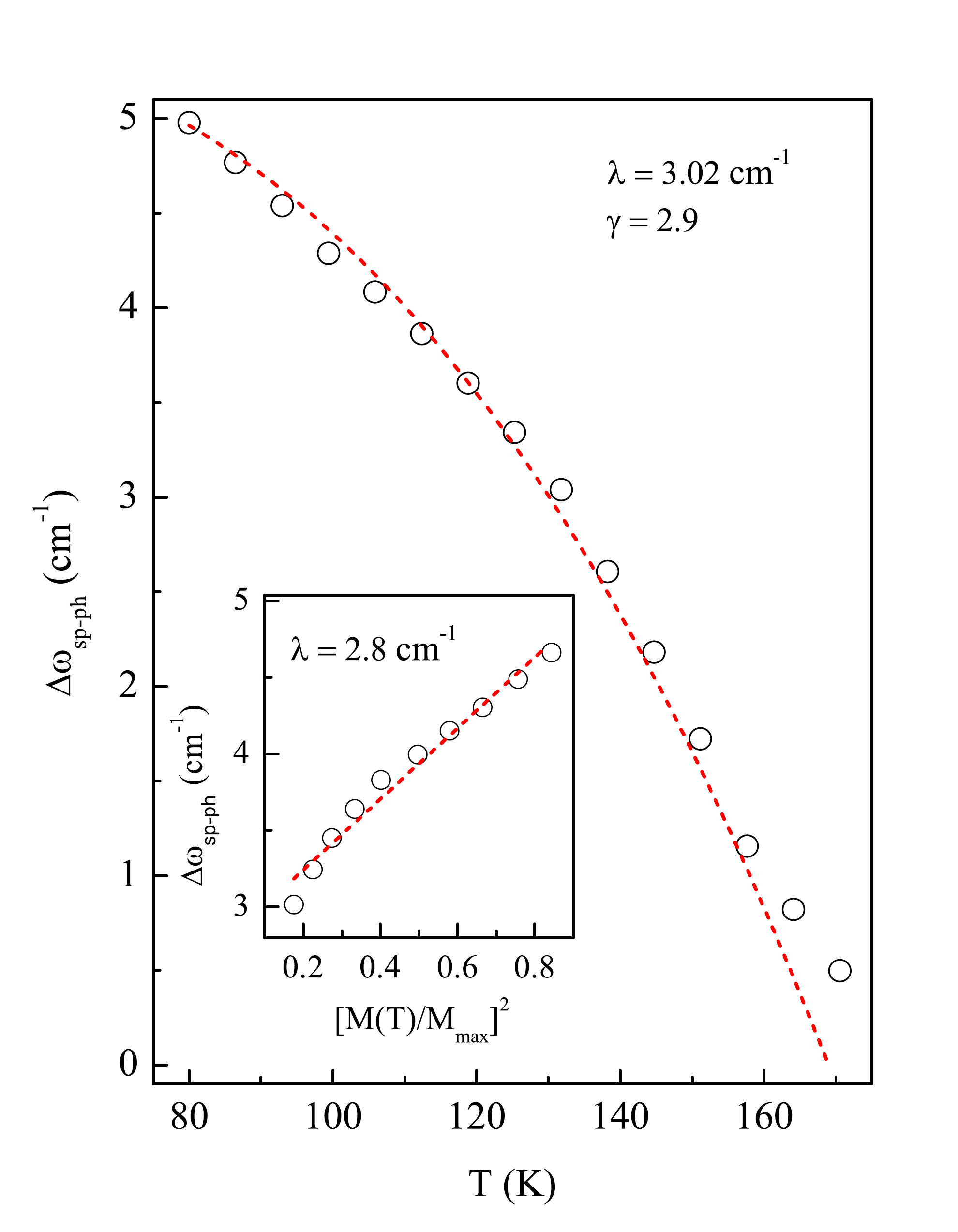}
 \caption{Temperature dependence of $\Delta\omega$ below $T_N$. The dotted line represents the fiiting using equation (2). Inset shows $\Delta\omega$ versus ($M(T)/M_{max})^2$ (circle) and its fitting (dotted line) using equation (3).}
 \end{figure}

Since different modes involve motion of different atoms, the associated coupling constant ($\lambda$) depends on how these motion change the bond lengths and bond angles involving the oxygen atoms which mediate magnetic exchange. As the antisymmetric stretching mode ($A_g$(7)) exhibits the largest deviation from the conventional anharmonic behavior below the transition (Figs. 5(b) and (e)), this should correspond to possibly the largest value of spin-phonon coupling.  FIG. 6 shows the thermal evolution of  $\Delta\omega_{sp-ph}$ below $T_N$ (circle) and its fitting with equation (3) (dotted line). The good fit obtained by considering only the symmetric Cr$^{3+}$-Cr$^{3+}$ interaction implies that the antisymmetric interaction (canted ferromagnetism) is very weak. We also found symmetric exchange coupling ($J_e$ = 11.06 K) is four times larger than the antisymmetric Dzyaloshinsky-Moriya interaction ($D$ = 2.64 K) by fitting the temperature dependence of magnetization with modified Curie-Weiss law given by Moriya \cite{moriya1960anisotropic,mcdannald2015magnetic, mahana2017complex}. From the fitting spin-phonon coupling constant ($\lambda$) of 3.02 cm$^{-1}$ and critical exponent ($\gamma$) of 2.9 are obtained, while the value calculated for the same from sub-lattice magnetization yields spin-phonon coupling constant ($\lambda$) of 2.8 cm$^{-1}$ which is in good agreement with that estimated from the order parameter. The obtained spin-phonon coupling in GdCrO$_3$ is quite comparable to the various systems estimated from Raman modes. For example, in antitiferromagnetic rutile structured MnF$_2$ and FeF$_2$ \cite{lockwood1988spin} the spin-phonon coupling strength for different modes are in the range from 0.4-1.3 cm$^{-1}$ and for Sr$_4$Ru$_3$O$_{10}$ \cite{gupta2006field}, $\lambda$ is  5.2 cm$^{-1}$. The above estimated coupling constant considers only  the nearest-neighbor Cr$^{3+}$-Cr$^{3+}$. In addition, there is an important contribution from Gd$^{3+}$-Cr$^{3+}$ interaction to spin-phonon coupling as discussed above. These results corroborate the existence of strong magneto-electric coupling in the system as evidenced from dielectric measurement as well as from enhancement of polarization with magnetic field \cite{bhadram2013spin, rajeswaran2012field}.

%%%%%%%%%%%%%%%%%%%%%%%%%%%%%%%%%%%%%%%%%%%%%%%%%%%%%%%%%%%%%%%%%

%%%%%%%%%%%%%%%%%%%%%%%%%%%%%%%%%%%%

\section{conclusion}
In conclusion, our experimental results clearly demonstrate that GdCrO$_3$ is locally non-centrosymmetric with $Pna2_1$ structure. Our calculations also support the observed ferroelectricity in GdCrO$_3$ through the determination of the detailed structure. There are competing structural instabilities in GdCrO$_3$ and the dominating one is of antiferrodistortive type and the weak polarization arises from the small ferroelectric instability resulting in Gd off-centering . The smaller displacement of Gd than that of Y leads to decrease in the strength of ferroelectricity in GdCrO$_3$ compared to YCrO$_3$, indicating a strong influence of $f$- electrons on the suppression of ferroelectric property of the system. Further, we found a large spin-phonon coupling of 3.02 cm$^{-1}$ from  antisymmetric stretching mode ($A_g$(7)) considering only the symmetric Cr$^{3+}$-Cr$^{3+}$ interaction, corroborating strong magneto-electric coupling in this material, which provides a complementary tool for the enhancement of ferroelectric polarization.
\section{Acknowledgment}
 B. R. would like to thank Prof. P. V. Satyam, IOP, Bhubaneswar for computational facilities and S. D. M. would like to thank IOP, Bhubaneswar, for kind hospitality. U. M. and D. T would like to thank ICTP-Elettra users program for the financial support to visit Elettra synchrotron centre, for experiments.
\bibliographystyle{apsrev4-1}
\bibliography{ref}

 \end{document}